%% file: main.tex
\documentclass[compsoc,conference,a4paper,10pt,times]{IEEEtran}
\input{preamble}

\makeatletter
\newcommand{\footnotewithoutnumber}[1]{%
  \begingroup
  \renewcommand{\thefootnote}{}%
  \renewcommand{\@makefntext}[1]{\noindent ##1}%
  \footnotetext{#1}%
  \endgroup
}
\makeatother

\begin{document}

\title{\codename: An Open Framework to Enable Arm CCA Research}

\author{\IEEEauthorblockN{
Andrin Bertschi
}
\IEEEauthorblockA{
\textit{ETH Zurich}
\\
Zürich, Switzerland \\
andrin.bertschi@inf.ethz.ch
}
\and
\IEEEauthorblockN{Shweta Shinde}
\IEEEauthorblockA{
\textit{ETH Zurich}
\\
Zürich, Switzerland \\
shweta.shinde@inf.ethz.ch
}
}

\maketitle

\input{sections/00_abstract}

\input{sections/10_introduction_v2}

\input{sections/20_background}

\input{sections/60_implementation}

\input{sections/70_evaluation}

\input{sections/80_casestudies}

\input{sections/90_conclusion}

\input{sections/95_ack}

\bibliographystyle{IEEEtran}
\bibliography{main}

\appendices
\crefalias{section}{appendix}

\input{sections/100_appendix}
\end{document}

%% file: preamble.tex
\usepackage{cite}
\usepackage{amsmath,amssymb,amsfonts}
\usepackage[table,xcdraw]{xcolor}
\usepackage{algorithmic}
\usepackage{graphicx}
\usepackage{textcomp}
\usepackage{bmpsize}
\usepackage{lipsum}
\usepackage{booktabs}
\usepackage{tabularx}
\usepackage{svg}
\usepackage[colorlinks=false,urlcolor=black,hidelinks]{hyperref}
\def\BibTeX{{\rm B\kern-.05em{\sc i\kern-.025em b}\kern-.08em
    T\kern-.1667em\lower.7ex\hbox{E}\kern-.125emX}}

\usepackage{listings}
\usepackage{xspace}
\usepackage{cleveref}
\usepackage{multirow}
\usepackage{enumitem}

\usepackage{adjustbox}
\usepackage{makecell}
\usepackage{pifont} %

\newcommand{\cmark}{\ding{51}} %
\newcommand{\xmark}{\ding{55}} %

\newcommand{\soc}{RK3588\xspace}
\newcommand{\codename}{{\sc OpenCCA}\xspace}

\newcommand{\tfa}{TF-A\xspace}
\newcommand{\rmm}{RMM\xspace}
\newcommand{\uboot}{U-Boot\xspace}

\newcommand{\new}[1]{#1}

\newcommand{\cvm}{CVM\xspace}
\newcommand{\cvms}{CVMs\xspace}

\newcommand{\RQone}{\hyperref[rq1]{\textbf{G1}}\xspace}

\newcommand{\RQthree}{\hyperref[rq3]{\textbf{G2}}\xspace}

\renewcommand{\paragraph}[1]{{\bf{\noindent #1}} }

\newcommand{\snapshot}[1]{ }

\crefname{figure}{Fig.}{Fig.}
\crefname{section}{Sec.}{Sec.}
\crefname{appendix}{Appx.}{Appx.}

\crefname{table}{Tab.}{Tab.}
\crefname{listing}{Lst.}{Lst.}

\newcommand{\spacehackfig}{\vspace{-1.5mm}}

\definecolor{mygreen}{RGB}{0, 128, 0}
\definecolor{myred}{RGB}{200, 0, 0}

\renewcommand{\cmark}{{\color{mygreen}\ding{51}}}
\renewcommand{\xmark}{{\color{myred}\ding{55}}}

%% file: sections/00_abstract.tex
\begin{abstract} 
Confidential computing has gained traction across major architectures 
with Intel TDX, AMD SEV-SNP, and Arm CCA. Unlike TDX and SEV-SNP, a key challenge in
researching Arm CCA is the absence of hardware support, forcing
researchers to develop ad-hoc prototypes on CCA emulators and non-CCA Arm
boards. This approach leads to high barriers to entry or duplicated efforts leading to unsound and inconsistent comparisons. To address this,
we present \codename, an open research platform that enables the
execution of CCA-bound code on commodity Armv8.2 hardware. By
systematically adapting the software stack (including bootloader,
firmware, hypervisor, and kernel), \codename
emulates CCA operations for performance evaluation while preserving
functional correctness. We demonstrate its effectiveness with typical
life-cycle measurements and case-studies inspired by prior CCA-based
papers on an easily available Arm v8.2 Rockchip board that costs $\$250$.

\end{abstract}

%% file: sections/10_introduction_v2.tex
\section{Introduction}

\footnotewithoutnumber{arXiv: This is the Systex2025 version of the paper.}

Intel and AMD, leaders in x86 platforms, enable confidential
computing with TDX and SEV-SNP respectively~\cite{tdx,sev-snp}. Arm, in 2021, 
announced Realm Management Extension (RME) which is an optional
feature on Armv9A to enable confidential computing
architecture (CCA)~\cite{cca}. Arm has rolled out support for building CCA-enabled
platforms, including emulator and software changes~\cite{arm-rmm,arm-fvp,arm-tfa}.
As witnessed by several recent works in top-tier venues, Arm CCA has already
received traction by providing a rich and fertile ecosystem for
innovative designs to improve confidential computing~\cite{shelter, cage, hitchhikerccs24, scrutinizer25, sok-commparison-study-tz-cca, acai, rcontainer, portal, via, xu2023virtcca, castes2024sharing, guarantee, cpc-usenix24, devlore-arxivv1, barriccade, aster, fortifypatch, cubevisor, twinvisor, asguard}. 

One fundamental hurdle in using Arm CCA is the lack of hardware
support, since no public Arm CPU supports CCA yet. 
\new{Arm's system emulator, the Fixed Virtual Platform (FVP)} provides functional correctness and instruction
counts. But, it is not cycle accurate which hinders any performance
measurements. To this end, researchers have resorted to building
best-effort performance prototypes where they {\em transplant} their
CCA-bound implementation to an Arm board that does not have CCA
support (e.g., Armv8) and replace the CCA instructions and
functionality with dummy operations. 
\begin{figure}[t]
\centering
\includegraphics[width=1\columnwidth]{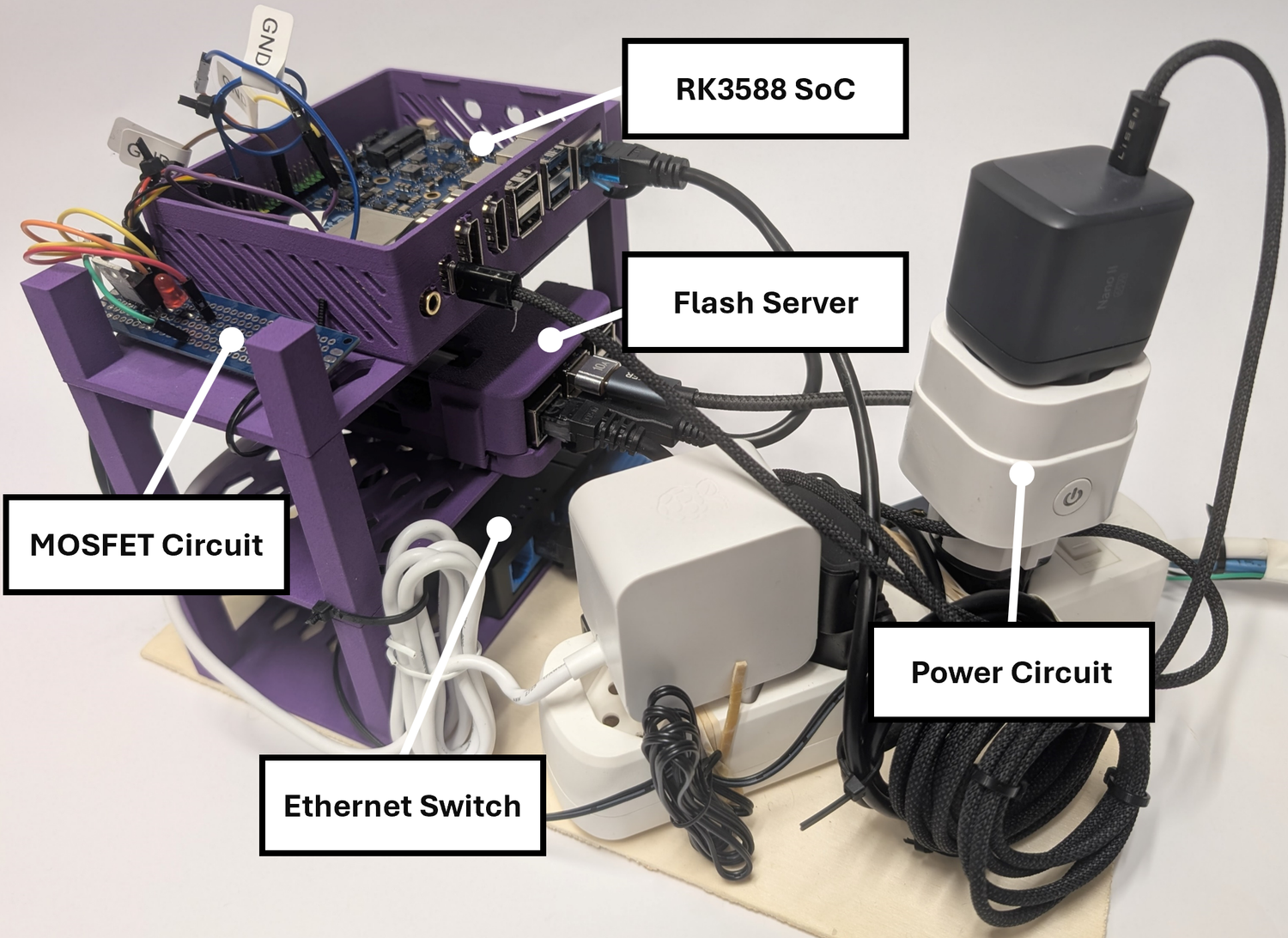} 
\caption{\codename tooling. The \soc connects over ethernet to a flash server (Raspberry Pi). It controls a MOSFET and power circuit to flash new firmware and exposes UART access.}
\label{fig:app:hw}
\spacehackfig
\end{figure}
Our survey shows that of the 19 papers released in the last 4 years, 15
spend significant effort on such transplantation (see~\Cref{tab:related:works}).
This not only creates unnecessary work for the authors, it also
detracts other researchers from experimenting on Arm CCA due to high
barrier for entry.
We observe that researchers choose different Arm boards for their
performance prototypes. 
These efforts are focused on the scope of the paper and do
not lend to full-fledged prototypes that are reusable by others. 
Thus researchers are repeating
the same efforts across boards, which is not the best use of their
time. 
The fragmentation in choice of boards further makes it
challenging to compare the performance across different papers, especially since some boards are 10 years old, cost $\$10,000$, or are no longer available (e.g., Juno R2). 
Worst of all, 4/19 papers omit hardware-based performance evaluation entirely due to lack of CCA support.

Our motivation to build \codename stems from: (a) our research experience prototyping early CCA works~\cite{acai,andrin-thesis}; (b) our roadblocks in subsequent performance prototypes~\cite{devlore-arxivv1,aster}; (c) personal communications with researchers in the community who want to benchmark performance for CCA-based defenses; and (d) large time gaps between Arm announcements and hardware rollouts (e.g., SEL2 announced in 2017, remains unavailable on commodity boards).
Taking inspiration from the {\em transplantation to performance
prototype} approach from prior efforts on CCA~\cite{shelter,hitchhikerccs24, scrutinizer25, cage, sok-commparison-study-tz-cca, acai,portal,rcontainer, castes2024sharing, via, xu2023virtcca, cubevisor, twinvisor, fortifypatch}, we build \codename (\cref{fig:app:hw}) that is
capable of executing CCA-bound code on a non-CCA Arm board. 
Specifically, we aim to enable lift-and-shift from Arm FVP to Arm board.
We first surveyed the most suitable boards for our goal (see \Cref{tab:hw:board:survey}) and chose the \soc Radxa Rock 5B due to its easy availability, support system, and
affordability.
We then systematically analyzed the entire stack from bootloader,
firmware, hypervisor, drivers, host and guest kernel to identify what aspects need to be adapted to emulate CCA
operations---for performance and
compatibility.
Finally, we added CCA awareness to all these components while
carefully selecting non-CCA operations that can best estimate CCA
overheads. 
\codename presents a standard development and measurement framework to evaluate CCA-based
solutions, akin to efforts on Intel SGX~\cite{opensgx} and RISC-V~\cite{keystone}.
Similar to OpenSGX~\cite{opensgx}, in its current form \codename does not aim to enforce CCA equivalent security on non-CCA boards. Lessons from ongoing efforts on  virtCCA~\cite{xu2023virtcca} can be coupled with \codename to address this limitation.

Our choice of a commodity board allows researchers to take their approach implemented on Arm FVP and lift-and-shift it to \codename for performance estimates and compatibility.
We demonstrate this by showcasing out-of-box use of \codename for reporting typical life-cycle metrics. Next, we run standard benchmarks on FVP and \codename to show that we preserve functionality.
Lastly, we then build two representative case-studies of CCA-based designs from prior works, under five hours each. 
\codename is available at~\cite{opencca}.

\begin{table}[]
    \caption{Survey of Work on CCA as of February 2025. Column 2-3 show if the work is implemented on Simulation Software and Arm Board respectively, with board architecture version in Column 4. Column 5-6 indicate if it is open-source (\cmark), closed (\xmark), or not applicable (N/A).   
    }
    \centering
    \renewcommand*{\arraystretch}{1.3}  
     \resizebox{1\columnwidth}{!}{%
    \begin{tabular}{lclccc}
        \toprule
        \textbf{Related Work} & \textbf{Sim.} & \textbf{Board} & \textbf{Arch.} & \textbf{FVP} & \textbf{Board}\\
        \midrule
        Cage~\cite{cage} & FVP & Juno R2 & 8.0 & \cmark & \cmark \\
        Shelter~\cite{shelter} & FVP & Juno R2 & 8.0 & \cmark & \xmark \\
        Scrutinizer~\cite{scrutinizer25} & FVP & Juno R2 & 8.0 & \cmark  & \xmark \\

        ACAI~\cite{acai} & FVP & Zynq UltraScale+ & 8.0 & \cmark & \xmark\\
        
        TZ \& CCA~\cite{sok-commparison-study-tz-cca} & FVP & Juno R2 & 8.0 & \xmark & \xmark\\
        
        HitchHiker~\cite{hitchhikerccs24} & FVP & Juno R2 & 8.0 & \xmark & \xmark \\
        
        FortifyPatch~\cite{fortifypatch} & FVP & Raspberry Pi 3B & 8.0 & \xmark & \xmark \\

        RContainer~\cite{rcontainer} & FVP & RK3399 Firefly  & 8.0 & \xmark & \xmark\\

        CubeVisor~\cite{cubevisor} & FVP & RK3399 Rock 4B & 8.0 & \xmark & \xmark \\

        TwinVisor~\cite{twinvisor} & FVP & HiSilicon Kirin 990 & 8.2 & \cmark & \xmark  \\

        Portal~\cite{portal} & FVP & OrangePi 5 Plus & 8.2 & \xmark & \xmark\\

        Des. \& Ver.~\cite{via} & FVP & Neoverse N1 & 8.2 &  \xmark & \xmark \\

        virtCCA~\cite{xu2023virtcca} & -- & Undisclosed & 8.4 & N/A & \xmark \\

        Sharing~\cite{castes2024sharing} & -- & AmpereOne & 8.6 & N/A & \xmark \\

         CPC~\cite{cpc-usenix24} & FVP & SEV SNP (x86) & -- & \xmark & \xmark \\

        GuaranTEE~\cite{guarantee} & FVP & -- & -- & \cmark & N/A \\

        Devlore~\cite{devlore-arxivv1} & FVP & -- & -- & \xmark & N/A\\
        BarriCCAde~\cite{barriccade} & QEMU & -- & -- & \xmark & N/A\\
        Aster~\cite{aster} & QEMU & -- & -- & \xmark & N/A\\

        \bottomrule
    \end{tabular}%
    }
     \label{tab:related:works}
\end{table}

%% file: sections/20_background.tex
\section{Arm CCA in a Nutshell}
\label{bg:armcca}

Prior to CCA, computation on Arm processors could execute in either normal or secure worlds. 
Arm CCA extends Arm's ISA with Realm Management Extensions (RME) to enable 2 new worlds: realm, and root (\cref{fig:opencca:bootflow}(a)). 
To isolate these worlds, RME adds Granule Protection Checks (GPCs) to each processing element (e.g., cores) which look up Granule Protection Tables (GPTs). 
The GPTs map physical addresses to their corresponding worlds and are programmed by the trusted firmware (TF-A) that executes in the root world (\cref{tab:ns_nse}).  
In addition to the worlds, the Arm ISA also allows computation to execute in one of the 4 exception levels (EL0-EL3). 
With CCA, only root world computation can execute in EL3. 

CCA enables the creation of confidential VMs (CVMs) in the realm world. 
To isolate mutually distrusting co-resident CVMs, CCA uses a trusted Realm Management Monitor (RMM) in realm EL2 which programs stage-2 translation tables for the CVMs.
The RMM exposes a Realm Management Interface (RMI) to the hypervisor and a Realm Service Interface (RSI) to the CVMs. 
The hypervisor invokes RMIs to create and manage the \cvms. 
Finally, the \rmm and the hypervisor use Secure Monitor Calls (SMCs) to communicate with \tfa.

%% file: sections/60_implementation.tex
\section{Design}
\label{sec:design}
We design \codename to meet the following goals:
\begin{itemize}[]
\item \RQone Enable Arm CCA on commodity Armv8 hardware
 with minimal software modifications and preserving functionality. \label{rq1} 

\item \RQthree Demonstrate adaptability
 and ease of integration as a research framework. \label{rq3}

\end{itemize}

\begin{figure}
\centering
\includegraphics[width=\columnwidth]{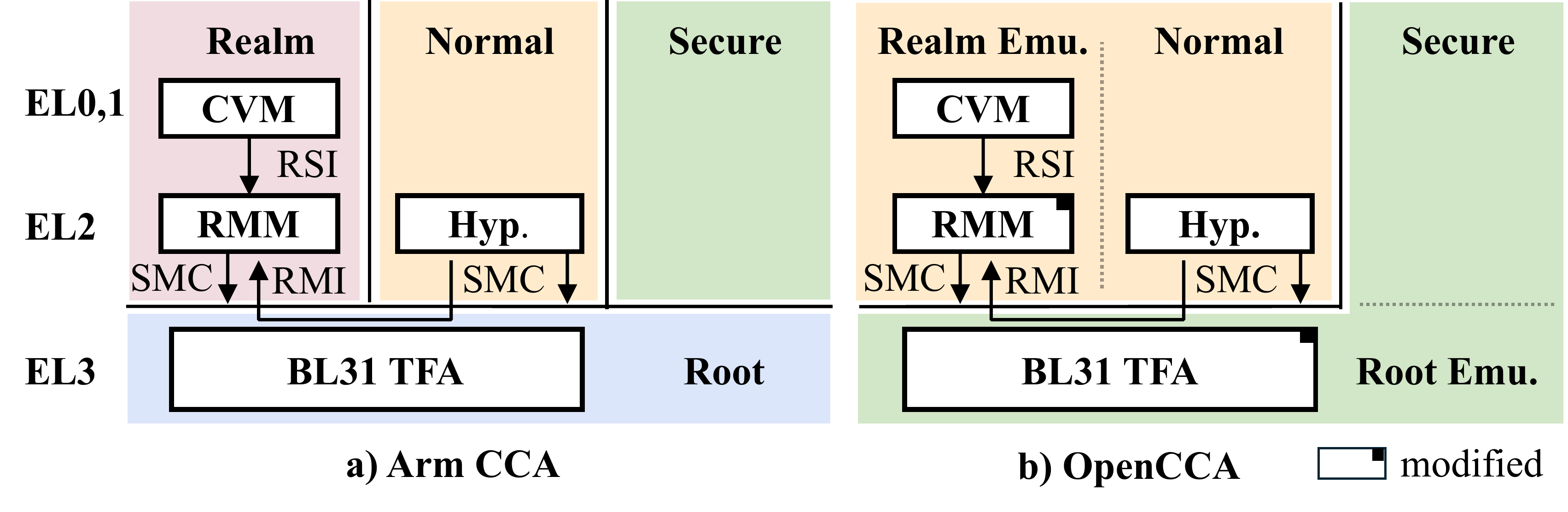}
   
     \caption{(a) Arm CCA. (b) \codename implements the realm world in the normal world (EL2-EL0) and root world in EL3 (i.e., secure world).
     }  
\label{fig:opencca:bootflow}
\end{figure}

\paragraph{Exploring Hardware Boards.}
In \Cref{tab:hw:board:survey} we list our survey on suitable boards for \codename.
We highlight the boards used by related works in gray color.
The Orange Pi 5 Plus board uses the same \soc as Radxa Rock 5b (\codename),
however, its software stack is currently more outdated.
RK3399, although cheaper than \soc, has older Cortex cores and,
as a result, fewer hardware features.
Interestingly, the Radxa Orion O6 board has modern cores and interrupt controller,
but currently lacks \tfa support and a publicly available technical reference manual
(TRM).
\new{In general, we consider a board suitable if two key conditions are met:
(1) EL3 must be flashable with custom firmware, i.e., not locked down by the vendor; and
(2) the board must offer at least rudimentary support of \tfa with source code available.
The \soc fulfills both. Furthermore, it is affordable, has good software support~\cite{collabora-rk3588} and a wide set of peripherals, making it a suitable platform for \codename at the time of writing.}

\paragraph{Main Insight.} 
We aim to lift-and-shift CCA-bound software tested on  Arm FVP to RK3588. 
We enlighten the three core firmware components on RK3588: the \tfa (\Cref{sec:design:tfa}), \rmm (\Cref{sec:design:rmm}), and \uboot (\Cref{sec:design:uboot}) such that they can emulate RME. 
This way, all other components such as hypervisor kernel, guest kernel, and the VMM remain unchanged---a developer can test it on Arm FVP with RME support and then transplant it to \codename. 
This design choice means that \codename remains independent of any particular hypervisor or \cvm configuration used (see \Cref{fig:opencca:bootflow}(b.)).

\input{sections/65_board_table.tex}

\subsection{Enabling \codename in \tfa}
\label{sec:design:tfa}

Since the \soc lacks the realm world, \codename first introduces a new world in software and addresses the absence of Granular Protection Table (GPT) instructions to enable \codename in \tfa.

\paragraph{Introducing a New World.} 
The Arm architecture does not automatically bank registers per world. 
Instead, the EL3 runtime firmware saves and restores the CPU context when switching between worlds. 
With RME, the security context of a core is defined by two bits in the Secure Configuration Register (\texttt{SCR\_EL3}), where the combination $\{\tt{NS},\tt{NSE}\}=11$ denotes the realm world (\Cref{tab:ns_nse}).
Without RME, the hardware lacks the NSE bit, making it impossible to distinguish the realm world at the architectural level.
During a world transition (e.g., SMC to schedule a \cvm), \tfa saves the CPU state in a memory region specific to the current core and world.
To switch worlds, \tfa updates the context to match the target world.
When EL3 exits, the CPU restores this context and continues execution in the new world.

To compensate for the missing NSE bit, \codename introduces a software bit: NSE' and stores it in the world context of each CPU.
Both NS (from \texttt{SCR\_EL3}) and NSE' (from memory) are then referenced to determine the active world.
This method allows for handling both synchronous (e.g., SMC) and asynchronous (e.g., interrupts) EL3 entries, but with added memory lookup.
By maintaining NS=1, \codename ensures that the realm world operates in the architectural normal world.
\new{Since we now multiplex both the realm and normal world within the architectural normal world, we must flush the EL2 TLB on each context switch between RMM and normal world hypervisor. 
This aligns with \RQone as we preserve functionality and keep the hypervisor agnostic of \codename}.
\new{We also maintain an alternative patchset (outside the scope of this evaluation) that eliminates the need for TLB flushes by requiring a small change in the hypervisor to reserve an address space identifier (ASID) range for the \rmm~\cite{opencca}.}

\begin{table}[]
    \caption{NS and NSE (\texttt{SCR\_EL3}) on Armv9 to select EL0/1/2 world. Root world is not encoded, always in EL3.}
    \centering
    \small
        \resizebox{0.7\columnwidth}{!}{%
    \begin{tabular}{lcclcc}
        \toprule
        \textbf{World} & \textbf{NS} & \textbf{NSE} & \textbf{World} & \textbf{NS} & \textbf{NSE} \\
        \cmidrule(lr{0.2em}){1-3} \cmidrule(lr{0.2em}){4-6} 
        Normal & 1 & 0 & Secure & 0 & 0 \\
        Realm   & 1 & 1 & Root   & - & - \\
        \bottomrule
    \end{tabular}
    \label{tab:ns_nse}%
    }
    \spacehackfig
\end{table}

\paragraph{Absence of GPT Instructions.} With the realm world in place, the challenge now shifts to memory isolation.
RME relies on GPTs to program world isolation, but \soc lacks the instructions to configure them.
Specifically, RME introduces the GPT base register (\texttt{GPTBR\_EL3}), GPC configuration register (\texttt{GPCCR\_EL3}), and TLB instructions (\texttt{TLBI PAALLOS}) which are unavailable.
To compensate, \codename replaces these with dummy system registers (\texttt{AFSRx}) and returns predefined values.
For instance, instead of querying the platform for GPC configuration, \codename returns a fixed configurable value.
We approximate the TLB instruction with a flush of the entire TLB cache (all shareability domains, exception levels, worlds).
This follows prior work~\cite{shelter, cage, scrutinizer25, hitchhikerccs24, acai, portal}.

\paragraph{Building the GPTs.}\label{sec:building:gpt} With GPT instructions substituted, \codename can now use them to initialize the protection tables.
The \soc does not utilize \tfa stage 2 bootloader (BL2) for early boot initialization.
Instead, it relies on \uboot to set up the platform.
Since \tfa implements GPT initialization in BL2 and BL2 is not deployed on the \soc, we integrate this functionality into BL31.
This allows \codename to account for GPT overhead during system boot without having GPC available on the board.

\subsection{Enabling \codename in \rmm}
\label{sec:design:rmm}
With \tfa modified for \codename, the system can now delegate memory between worlds and execute a bare-metal payload in realm EL2.
The next step is to enable \codename in the \rmm to run \cvms on the \soc.

\paragraph{No Small Translation Tables.} The \rmm uses identity mappings in low virtual memory (TTBR0) to map data and code that is shared across cores.
It places core-private memory in high virtual memory (TTBR1).
For TTBR1, the \rmm only requires an address space size of 2 MB.
As a result, it uses Small Translation Tables (TTST) for high-memory mappings.
TTST is a feature introduced in Armv8.4 and decreases the lower limit on the size of translation tables~\cite{arm-features}.
As such, the page table walk is shorter because the MMU traverses fewer levels to reach the leaf node. 
The \soc lacks this feature. Hence, we change TTBR1 mappings to use an address space size of 64 MB. This forces a base level of 2, the smallest level supported on the \soc.
With the \rmm's own memory mappings correctly configured, \codename can now address \cvm stage-2 mappings.

\paragraph{Stage-2 Mappings without FWB.}
Force Write Back (FWB) is an Armv8.3 feature that enables the hypervisor to enforce write-back behavior on non-cacheable translation mappings set by the VM~\cite{arm-features}. This eliminates the need for explicit cache maintenance because guest writes become immediately visible to the hypervisor.
In the \rmm, stage-2 management requires FWB; however, \soc does not implement this feature.
As a result, we encountered issues during the \cvm early boot process, leading to stale data and inconsistent crashes.
To mitigate this, \codename changes the memory attributes in stage-2 and adds cache maintenance instructions.

\paragraph{Timer Virtualization.} Armv8.6 introduces new system registers to control time for VMs with Enhanced Counter Virtualization (ECV)~\cite{arm-features}. This functionality is missing on the \soc.
This includes an offset for time (\texttt{CNTPOFF}) between the guest and the hypervisor, which can be useful in scenarios like live migration, where counters may differ between source and destination hosts.
The \rmm specification mandates that these offsets must be fixed for the lifetime of a realm~\cite{rmm-spec}. Since we do not have ECV available, \codename sets these offsets to zero.

The \rmm also controls the firing of timer interrupts;
when a physical timer occurs, the \rmm traps the \cvm into EL2 and transitions control to the normal world hypervisor.
Subsequently, the hypervisor delivers a virtual interrupt to the \cvm. This design keeps the \rmm's TCB minimal and delegates interrupt management to the hypervisor.
The \rmm uses EL2 system registers to mask the physical timer, preventing an immediate exit of the \cvm upon re-entry (\texttt{CNTPMASK}).
Due to missing RME and ECV, this mechanism is unavailable and leads to \cvm stalls on the \soc.
\codename addresses this by overriding timer masking with EL0 registers, ensuring the guest can continue making progress (\texttt{CNTP\_CTL\_EL0}).

\paragraph{FP Traps and Timer Interplay.} Until this point, the \cvm successfully boots to EL0, but experiences stalls due to traps caused by lazy floating-point (FP) state restoration.
As an optimization, the \rmm only restores FP registers when the \cvm actually uses them.
Specifically, the \rmm traps on the first FP used, restores the FP state and then disables further traps until the next transition to the hypervisor.
This leads to a loop where the \cvm traps for FP state restoration, the \rmm restores the FP state, but before execution can proceed, a physical timer interrupt forces a transition to the hypervisor, repeating the cycle.
To prevent this, \codename keeps the timer masked if the previous \cvm exit was triggered by FP state restoration.

\subsection{Enabling \codename in \uboot}
\label{sec:design:uboot}

At this stage, \codename successfully boots a \cvm in realm EL1/EL0. Bundling the \rmm into the firmware image, although presented last, is the first step during compilation and a prerequisite for the boot process.

\paragraph{Bundling a new Firmware.} \uboot uses Binman~\cite{uboot-binman} to package multiple firmware components into a single image. We introduce the \rmm as a new firmware component and ensure that the firmware chain includes both \tfa and the \rmm.

\section{Implementation}
We prototype \codename on Arm's reference implementation of CCA.
\Cref{tab:firmware-tcb} summarizes our firmware stack and lines of code (LoC) modified to support \codename.
\new{\codename introduces minimal; 940 (+0.3\%), 1440 (+6\%) and 216 (+0.01\%) LoC for \tfa, \rmm and \uboot respectively (\RQone), compared to the total sizes of 309K, 25K and 1.5M LoC (C/C++/ASM).}
We keep hypervisor, \cvm, and VMM (kvmtool) unchanged.
We see that platform-specific code constitutes a significant portion of the overall code changes.
This includes a new console driver, a new memory layout for the \soc, and GPT code we moved from BL2 in \tfa to BL31.
\new{We refer to our extended version for more implementation details~\cite{opencca-arxiv}.}

\begin{table}
    \centering
    \caption{System stack in \codename. Firmware based on latest versions available at the outset of project. No changes in software stack.}
    \label{tab:firmware-tcb}
    \renewcommand{\arraystretch}{1.1}
    \setlength{\tabcolsep}{4pt}
    \resizebox{1\columnwidth}{!}{%
    \begin{tabular}{llrrll}
        \toprule
        \multicolumn{4}{c}{\textbf{Firmware Stack}} & \multicolumn{2}{c}{\textbf{Software Stack}} \\
        \cmidrule(lr{0.2em}){1-4} \cmidrule(lr{0.2em}){5-6} 
        \textbf{Component} & \textbf{Version} & \textbf{Modification} & \multicolumn{1}{c}{\makecell{\textbf{RK3588} \\ \textbf{Specific}}} & \textbf{Component} & \textbf{Version} \\
        \midrule
        TF-A~\cite{version-tfa} & v2.11     & 940 LoC & 59\% 
        & Linux Hyp.~\cite{version-cca-patchset-host, version-linux-host, collabora-rk3588} & v6.12.0 \\
        
        TF-RMM~\cite{version-rmm}  & v0.5.0 & 1440 LoC & 16\% & Linux \cvm \cite{version-cca-patchset-host, version-linux-host} & v6.12.0\\
        U-Boot~\cite{version-uboot}   & v2024.01& 216 LoC & 0\% & kvmtool \cite{version-kvmtool} &  v3/cca  \\
        \bottomrule
    \end{tabular}%
    }
    \spacehackfig
\end{table}

%% file: sections/65_board_table.tex
\begin{table*}[h]
\caption{Exploring Hardware Boards for \codename. \tfa Code: \tfa ported to board and source code publicly available. Gray Highlighted: Boards used in related works in \Cref{tab:related:works}. Green and Bold: Board used in \codename.}
    \centering
    \label{tab:hw:board:survey}
        \renewcommand*{\arraystretch}{1.1}    
\resizebox{0.8\textwidth}{!}{%
\begin{tabular}{@{}lrllrllc@{}}
\toprule
\textbf{Board}                           & \textbf{Released} & \textbf{SoC}              & \textbf{GIC}      & \textbf{Price (USD)}    & \textbf{Cores}           & \textbf{GPU}             & \multicolumn{1}{r}{\textbf{TF-A Code}} \\ \midrule

Intel Stratix 10 SX DK          & 2013     & Intel Stratix 10 & GICv2    & 9,000          & A53             & N/A             & \cmark          \\
AmloGIC Meson S905 (GXBB)       & 2015     & S905             & GICv2    & unknown        & A53             & Mali 450        & \cmark                    \\
HiKey                           & 2015     & Kirin 620        & GICv2    & 75-100         & A53             & Mali 450        & \cmark          \\

\rowcolor[HTML]{C0C0C0} 
Arm Juno r2                     & ca. 2015 & Juno r2 SoC      & GICv2    & 10,000         & A72, A53        & Mali T624       & \cmark          \\

NXP i.MX7 WaRP7                 & 2016     & i.MX 7 Solo      & GICv2    & 100            & A7, M4          & N/A             & \cmark          \\
A64-OLinuXino                   & 2016     & Allwinner A64    & GICv2    & 100            & A53             & Mali 400        & \cmark                     \\
AmloGIC Meson S905x (GXL)       & 2016     & S905x            & GICv2    & unknown        & A53             & Mali 450 MP3    & \cmark                     \\
NXP i.MX 8QM MEK                & 2016/17  & i.MX 8QM         & GICv3    & 1,200          & A72, A53, M4F   & GC7000XSVX      & \cmark          \\
NXP i.MX 8MQ EVK                & 2016/17  & i.MX 8MQ         & GICv3    & 500            & A53, M4         & GC7000Lite      & \cmark          \\
NXP i.MX 8ULP EVK               & 2016/17  & i.MX 8 ULP       & GICv3    & 550-650        & A53, M33        & GC520           & \cmark          \\

\rowcolor[HTML]{C0C0C0} 
Xilinx Zynq ZCU102 EVK          & ca. 2017 & 2FFVB1156E       & GICv2    & 3,200           & A53, R5F        & Mali 400 RP2    & \cmark          \\

AmloGIC Meson A113D (AXG)       & 2017     & S400             & GICv2    & unknown        & A53             & 2D GFX Engine   & \cmark                    \\
HiKey 960                       & 2017     & Kirin 960 SoC    & GICv2    & 250            & A73, A53        & Mali G71 MP8    & \cmark          \\
AmloGIC Meson S905X2 (G12A)     & 2018     & S905x2           & GICv2    & unknown        & A53             & Mali-G31 MP2    & \cmark                    \\
HiKey 970                       & 2018     & Kirin 970        & GICv2    & 300            & A73, A53        & Mali G72 MP12   & \cmark          \\
\rowcolor[HTML]{C0C0C0} 
Raspberry Pi 3 (B+)             & 2018     & BCM2837B0        & custom   & 25             & A53             & VideoCore IV    & \cmark          \\
Intel Agilex 7M HBM2e DK        & 2019     & Intel Agilex 7   & GICv2    & 10,000         & A53             & N/A             & \cmark          \\
Marvell CEx7 CN9132 EVB         & 2019     & CN9132           & GICv2    & 600-700        & A72             & N/A             & \cmark          \\
Ziver MTK8183 Dev. Board        & 2019     & MT8183           & GICv3    & 150            & A73, A53        & Mali G72 MP3    & \cmark                    \\
Raspberry Pi 4                  & 2019     & BCM2711          & GICv2    & 35             & A72             & VideoCore VI    & \cmark          \\

\rowcolor[HTML]{C0C0C0} 
Huawei Mate 30 Pro &  2019    & Kirin 990 & unknown    & 300      & A76, A55  & Mali-G76 &  \xmark   \\

\rowcolor[HTML]{C0C0C0} 
Arm Neoverse N1 SDP             & 2020     & Dawn Ares        & GICv4.1  & 10,000         & N1 & HDLCD           & \cmark          \\

Aspeed AST2700 EVB              & 2020     & AST2700          & GICv3    & unknown        & A35, M4         & AST2700 2D VE   & \cmark  \\

\rowcolor[HTML]{C0C0C0} 
RK3399 Rock4                    & 2021     & RK3399           & GICv3    & \textless{}200 & A72, A53        & Mali-T864       & \cmark          \\

NVIDIA Jetson TX2 NX DK         & 2021     & Tegra X2         & GICv2    & 350            & Denver2, A57    & GP10B           & \cmark                     \\
MediaTek 8186                   & 2021     & Kompanio 520     & GICv3    & unknown        & A76, A55        & Mali-G52 MP2    & \cmark          \\
MediaTek 8192                   & 2021     & Kompanio 820     & GICv3    & unknown        & A55, A76        & Mali G57 MC5    & \cmark          \\
MediaTek 8188                   & 2022     & Kompanio 838     & GICv3    & unknown        & A55, A78        & Mali G57 MC3    & \cmark          \\
MediaTek 8195                   & 2022     & Kompanio 1380    & GICv3    & unknown        & A55, A78        & Mali G57 MC5    & \cmark          \\

Genio 700 (MT8390)              & 2023     & MT8390           & GICv3    & 700            & A78, A55        & Mali-G57        & \cmark          \\

\rowcolor[HTML]{C0C0C0} 
Orange Pi 5 Plus                & 2023     & RK3588           & GICv3    & \textless{}200 & A76, A55        & Mali-G610       & \cmark          \\

\rowcolor[HTML]{C0C0C0} 
Supermicro MegaDC (Server)                & 2023     & AmpereOne           & unknown & unknown & custom built      & N/A       & \xmark          \\

Raspberry Pi 5                  & 2023     & BCM2712          & GICv2    & 120            & A76             & VideoCore VII   & \cmark          \\

\rowcolor[HTML]{D2E6C8} 
\textbf{Radxa Rock 5b (\codename)} & \textbf{2023}     & \textbf{RK3588}        & \textbf{GICv3}    & \textbf{250}            & \textbf{A76, A55}        & \textbf{Mali-G610}       & \cmark          \\

NXP i.MX 93 QS EVK              & ca. 2024 & i.MX 93          & GICv3/v4 & 300            & A55, M33        & N/A             & \cmark                    \\

Arrow AXE5-Eagle DK             & 2024     & Intel Agilex 5   & GICv3    & 900-1,000      & A55, A76        & N/A             & \cmark          \\

Radxa Orion O6                  & 2024     & Cix CD8180       & GICv4    & 500-600        & A720, A520      & Imtls. G720 MC6 & \xmark                    \\

\bottomrule
\end{tabular}
}
\spacehackfig
\end{table*}

%% file: sections/70_evaluation.tex
\section{Evaluation}
We demonstrate \codename functionality and report runtime measurements.

\paragraph{Experimental Setup.} 
We boot Linux in a \cvm with 1 vCPU, and 256MB and 1GB of RAM. We pin kvmtool to core 2 and isolate the core from general-purpose scheduling ($\tt{isolcpus}$). The normal world hypervisor uses the 4 Cortex A55 cores on the \soc (see \cref{tab:board-specs}) and the CPU governor userspace, that we set to a fixed frequency of 1.8GHz.
For instructions and cycles on the \soc, we use Performance Monitor Unit (PMU) with events: Instructions Retired and Cycles.
\new{We build \tfa and \rmm in release mode and disable all but ERROR output.}

\paragraph{PMU across Worlds.} \tfa and \rmm manage performance counters by saving and restoring them for EL2 and EL1 upon context switch.
However, for \codename's evaluation, we are interested in measuring overhead across all exception levels.
To achieve this, we introduce a patch set that bypasses the standard save-and-restore mechanism for PMU counters across worlds and exception levels.

\paragraph{Verifying Results with FVP.}
For completeness, we benchmark a CPU-intensive workload on both RK3588 and FVP using identical binaries (hypervisor, CVM, payload), and manually verify that the results are consistent.

\paragraph{Benchmarks.}
To benchmark \codename, we review related works~\cite{acai, shelter, cage, portal} and identify key performance metrics they use to evaluate overheads on Arm CCA. These include RMI and context switch costs, VM boot overheads, and GPT costs across different setups (i.e., a baseline against new changes introduced by research).
\new{We report cycles and instructions with standard deviation and average \cvm boots across 100 iterations and SMC/RMI benchmarks across 5 million invocations.}

In \Cref{tab:benchmark_results}, we show an overview of \codename runtime measurements.
As expected, \cvm boot increases with larger RAM sizes due to the additional memory delegation required. 
We further compare \codename against a \textit{Two-GPT} case study (see \Cref{sec:casestudies}) and observe an overhead of 1.19\% in instructions and 1.15\% in cycles when booting a \cvm with 1GB of RAM.
\new{For context switch costs, an SMC round trip that saves and restores the world context without invoking a service in \tfa incurs 182 instructions and 421 cycles.}
\new{Similarly, an RMI round trip that directly returns from the \rmm requires 932 instructions and 3370 cycles, which is 213 cycles less than an RMI that queries the version.}
\new{Microarchitectural noise affects short calls; grouping multiple calls into a batch may reduce variance.}

\begin{table}[]
    \centering
    \caption{Radxa Rock5b Board Specifications.}
    \renewcommand{\arraystretch}{1.1}
    \resizebox{1\columnwidth}{!}{%
    \begin{tabular}{ll ll}
        \toprule
        \textbf{Component} & \textbf{Specification} & \textbf{Component} & \textbf{Specification} \\
        \midrule
        \textbf{SoC} & Rockchip RK3588 & \textbf{GPU}  & Mali-G610  \\
        \textbf{Board} & \multirow{2}{*}{\shortstack[l]{Radxa Rock5b \\ v1.46-2023-11.06}} & \textbf{RAM} & 16 GB \\
        & & \textbf{Storage} & 64 GB eMMC \\
      
        \textbf{CPU} & \multirow{2}{*}{\shortstack{4× Cortex-A76 @ 2.4 GHz \\ 4× Cortex-A55 @ 1.8 GHz}} & \textbf{IRQ} & GICv3 \\
        & & \textbf{PCI} & PCIe 3.0, 2.0 \\
        \bottomrule
    \end{tabular}%
    }
    \label{tab:board-specs}
    \spacehackfig
\end{table}

%% file: sections/80_casestudies.tex
\section{Case Studies}
\label{sec:casestudies}
We address \RQthree and show \codename's adaptability by reimplementing prior designs, reporting implementation time and modified LoC.

\paragraph{Two-GPT.}
In this case study, we implement a dual GPT mechanism building on designs proposed in \cite{acai, cage, portal, shelter, devlore-arxivv1, aster, hitchhikerccs24, scrutinizer25, rcontainer, cubevisor, fortifypatch}. We introduce a second GPT while retaining the existing GPT.
During system boot, we mark all memory in GPT2 as root world.
Upon memory delegation (\texttt{RMI\_GRANULE\_DELEGATE}), GPT1 marks the memory as realm world, while GPT2 marks it as normal world. During memory undelegation (\texttt{RMI\_UNGRANULE\_DELEGATE}), the process is reversed, restoring normal world in GPT1 and root world in GPT2. We modify \codename memory layout to reserve space for both GPT data structures. We use the existing GPT1 code as a template and duplicate it for GPT2.
In total, we change 7 files in \tfa (2348 lines added, 13 deleted), and the implementation took 4 person hours.
Delegating a single page with Two-GPT takes $3488$ (+21.7\%) instructions and $8654$ (+8.3\%) cycles compared to delegation with a single GPT. 

\paragraph{Shadow GPT.}
We implement a GPT management design that Shelter~\cite{shelter} uses for Shelter-Apps. For GPT construction, Shelter creates a shadow GPT, a pre-configured template that is copied instead of being built from scratch. We modify \soc memory layout to reserve space for new GPTs. In total, we change 10 files in \tfa (1398 lines added, 24 deleted) and the implementation took 5 person hours. \new{Creating a shadow GPT takes $50.86$M instructions and $34.61$M cycles.}

\begin{table}[]
    \centering
    \caption{Evaluation, RT: Round Trip, Delegate: 4KB}
    \small
    \setlength{\tabcolsep}{4pt}
    \renewcommand{\arraystretch}{1}
    \resizebox{0.8\columnwidth}{!}{%
    \begin{tabular}{lrrrrc}
        \toprule
        \multirow{2}{*}{\textbf{Benchmark}} 
        & \multicolumn{2}{c}{\textbf{Mean}} 
        & \multicolumn{2}{c}{\textbf{Stdev}} 
        & \multirow{2}{*}{\textbf{Scale}} \\ \cmidrule(lr{0.2em}){2-3} \cmidrule(lr{0.2em}){4-5} 
        & \textbf{Instr} & \textbf{Cycles} 
        & \textbf{Instr} & \textbf{Cycles} 
        & \\
        \midrule
        \multicolumn{6}{c}{\textbf{\codename}} \\
        \midrule
        \cvm Boot 256 MB            & 1900 & 2647 & 6  & 15  & 1M \\
        \cvm Boot 1 GB              & 2015 & 2869 & 8  & 18  & 1M \\
        \midrule
        RMI Delegate                & 2865 & 7988 & 187 & 365 & 1 \\
        RMI Version                 & 994  & 3583 & 120 & 222 & 1 \\
        RMI RT                      & 932  & 3370 & 115 & 209 & 1 \\
        SMC RT                      & 182  & 421  & 44  & 68  & 1 \\
        \midrule
        \multicolumn{6}{c}{\textit{Two-GPT} Case Study} \\
        \midrule
        \cvm Boot 256 MB            & 1928 & 2690 & 9  & 10  & 1M \\
        \cvm Boot 1 GB              & 2039 & 2902 & 7  & 18  & 1M \\
        \midrule
        RMI Delegate                & 3488 & 8654 & 182 & 372 & 1 \\
        \bottomrule
    \end{tabular}%
    }
    \label{tab:benchmark_results}
    \spacehackfig
\end{table}

%% file: sections/90_conclusion.tex
\section{Using \codename for Research}
\label{sec:rec:usage}
\new{Prior work uses FVP's instruction counts as a proxy for performance (\cref{tab:related:works}). }
It is not a meaningful metric; the FVP is not designed for timing analysis but only functional validation, it lacks realistic models for out-of-order execution, superscalar pipelines, and memory hierarchies~\cite{fvp-not-cycle-accurate}.
In the absence of CCA-enabled hardware, \codename provides a best-effort estimation.
We argue that this approach introduces fewer errors than relying on the FVP since the impact of missing RME features (e.g., GPC, realm/root world, memory encryption) is smaller than the inaccuracies caused by the lack of a microarchitectural performance model.
Therefore, we recommend using a hardware prototype for performance evaluation and limiting the FVP to functional validation.

\paragraph{Leveraging Features on \soc.} 
Researchers can modify \codename \tfa and \rmm to use existing hardware functionality (e.g., cryptography extensions, PCIe, and SMMU integrate into \tfa as \soc natively supports them).
\new{We recommend identifying the feature in the A-profile list~\cite{arm-features}, confirming it is in Armv8.2, and verifying its availability in the TRM for \soc.}

\paragraph{Addressing Missing Features on \soc.}
If the \soc lacks a hardware feature (e.g., Memory Tagging, PAC), we propose two solutions: simulate functionality in software and approximate overheads like we do for RME, or upgrade to a newer board.
For \codename, a new hardware target with more hardware features requires only a subset of the existing porting work, \new{as more CCA-related functionality is already natively supported by the hardware.}
To effectively diagnose issues, we suggest a platform with hardware debugging support~\cite{opencca-arxiv}.

\section{Conclusion}
\vspace{-5pt}

Researching on Arm CCA remains challenging due
to the lack of hardware support, causing
inefficiencies and inconsistent performance comparisons. 
To overcome this, we introduce \codename, an open research framework
that enables CCA-bound code execution on affordable Armv8.2 hardware.
\codename emulates CCA operations for performance evaluation while maintaining functionality and enabling lift-and-shift from FVP.
\newline

%% file: sections/95_ack.tex
\vspace{-5pt}
\paragraph{Acknowledgment.}
We thank the anonymous reviewers, Supraja Sridhara, and Mark Kuhne for their constructive feedback. Thanks to Dual Tachyon for support with RK3588  debugging.

%% file: main.bbl
\begin{thebibliography}{10}
\providecommand{\url}[1]{#1}
\csname url@samestyle\endcsname
\providecommand{\newblock}{\relax}
\providecommand{\bibinfo}[2]{#2}
\providecommand{\BIBentrySTDinterwordspacing}{\spaceskip=0pt\relax}
\providecommand{\BIBentryALTinterwordstretchfactor}{4}
\providecommand{\BIBentryALTinterwordspacing}{\spaceskip=\fontdimen2\font plus
\BIBentryALTinterwordstretchfactor\fontdimen3\font minus
  \fontdimen4\font\relax}
\providecommand{\BIBforeignlanguage}[2]{{%
\expandafter\ifx\csname l@#1\endcsname\relax
\typeout{** WARNING: IEEEtran.bst: No hyphenation pattern has been}%
\typeout{** loaded for the language `#1'. Using the pattern for}%
\typeout{** the default language instead.}%
\else
\language=\csname l@#1\endcsname
\fi
#2}}
\providecommand{\BIBdecl}{\relax}
\BIBdecl

\bibitem{tdx}
Intel,
  ``{\href{https://www.intel.com/content/www/us/en/developer/articles/technical/intel-trust-domain-extensions.html}{Intel
  Trust Domain Extensions (Intel TDX)}},'' {Accessed: Sep.~2, 2024}.

\bibitem{sev-snp}
AMD,
  ``\href{https://www.amd.com/content/dam/amd/en/documents/epyc-business-docs/white-papers/SEV-SNP-strengthening-vm-isolation-with-integrity-protection-and-more.pdf}{AMD
  SEV-SNP\: Strengthening VM Isolation with Integrity protection and more},''
  {Accessed: Sep.~2, 2024}.

\bibitem{cca}
ARM,
  ``\href{https://www.arm.com/why-arm/architecture/security-features/arm-confidential-compute-architecture}{Arm
  Confidential Compute Architecture ({ARM-CCA})},'' {Accessed: Jan.~1, 2025}.

\bibitem{arm-rmm}
------, ``\href{https://www.trustedfirmware.org/projects/tf-rmm}{Trusted
  Firmware Implementation of the Realm Management Monitor (RMM), Project
  Page},'' {Accessed: Sep.~2, 2024 }.

\bibitem{arm-fvp}
------,
  ``\href{https://developer.arm.com/documentation/100966/1121/?lang=en}{Fast
  Models Fixed Virtual Platforms (FVP), Reference Guide, Version 11.21},''
  {Accessed: Jan.~1, 2025,}.

\bibitem{arm-tfa}
------, ``\href{https://trustedfirmware-a.readthedocs.io/en/latest/}{Arm
  Trusted Firmware-A, Project Page},'' {Accessed: Sep.~1, 2024, }.

\bibitem{shelter}
Y.~Zhang, Y.~Hu, Z.~Ning, F.~Zhang, X.~Luo, H.~Huang, S.~Yan, and Z.~He,
  ``{SHELTER}: Extending arm {CCA} with isolation in user space,'' in
  \emph{USENIX Security}, 2023.

\bibitem{cage}
C.~Wang, F.~Zhang, Y.~Deng, K.~Leach, J.~Cao, Z.~Ning, S.~Yan, and Z.~He,
  ``Cage: Complementing arm cca with gpu extensions,'' in \emph{NDSS}, 2024.

\bibitem{hitchhikerccs24}
C.~Zhang, J.~Zeng, Y.~Zhang, A.~Ahmad, F.~Zhang, H.~Jin, and Z.~Liang, ``{The
  HitchHiker's Guide to High-Assurance System Observability Protection with
  Efficient Permission Switches},'' in \emph{{ACM CCS}}, 2024.

\bibitem{scrutinizer25}
Y.~Zhang, F.~Zhang, X.~Luo, R.~Hou, X.~Ding, Z.~Liang, S.~Yan, T.~Wei, and
  Z.~He, ``{SCRUTINIZER: Towards Secure Forensics on Compromised TrustZone},''
  in \emph{NDSS}, 2025.

\bibitem{sok-commparison-study-tz-cca}
H.~Huang, F.~Zhang, S.~Yan, T.~Wei, and Z.~He, ``{SoK: A Comparison Study of
  Arm TrustZone and CCA},'' in \emph{2024 International Symposium on Secure and
  Private Execution Environment Design (SEED)}, 2024.

\bibitem{acai}
S.~Sridhara, A.~Bertschi, B.~Schlüter, M.~Kuhne, A.~Aliberti, and S.~Shinde,
  ``{Acai: Protecting Accelerator Execution with Arm Confidential Computing
  Architecture},'' in \emph{USENIX Security}, 2024.

\bibitem{rcontainer}
Q.~Zhou, W.~Cao, X.~Jia, P.~Liu, S.~Zhang, J.~Chen, S.~Xu, and Z.~Song,
  ``{RContainer: A Secure Container Architecture through Extending ARM CCA
  Hardware Primitives},'' in \emph{NDSS}, 2025.

\bibitem{portal}
F.~Sang, J.~Lee, X.~Zhang, and T.~Kim, ``{PORTAL: Fast and Secure Device Access
  with Arm CCA for Modern Arm Mobile System-on-Chips (SoCs)},'' in \emph{IEEE
  S\&P}, 2025.

\bibitem{via}
X.~Li, X.~Li, C.~Dall, R.~Gu, J.~Nieh, Y.~Sait, and G.~Stockwell, ``{Design and
  Verification of the Arm Confidential Compute Architecture},'' in \emph{USENIX
  OSDI}, 2022.

\bibitem{xu2023virtcca}
X.~Xu, W.~Wang, Y.~Wu, C.~Wang, H.~Zhu, H.~Ma, Z.~Min, Z.~Pang, R.~Hou, and
  Y.~Jin, ``{virtCCA: Virtualized Arm Confidential Compute Architecture with
  TrustZone},'' \emph{arXiv preprint arXiv:2306.11011}, 2023.

\bibitem{castes2024sharing}
C.~Castes and A.~Baumann, ``{Sharing is leaking: blocking transient-execution
  attacks with core-gapped confidential VMs},'' in \emph{ACM ASPLOS}, 2024.

\bibitem{guarantee}
S.~Siby, S.~Abdollahi, M.~Maheri, M.~Kogias, and H.~Haddadi, ``{GuaranTEE:
  Towards Attestable and Private ML with CCA},'' in \emph{Proceedings of the
  4th Workshop on Machine Learning and Systems}, 2024.

\bibitem{cpc-usenix24}
J.~Chen, Z.~Mi, Y.~Xia, H.~Guan, and H.~Chen, ``{CPC}: Flexible, secure, and
  efficient {CVM} maintenance with confidential procedure calls,'' in
  \emph{USENIX ATC}, 2024.

\bibitem{devlore-arxivv1}
A.~Bertschi, S.~Sridhara, F.~Groschupp, M.~Kuhne, B.~Schlüter, C.~Thorens,
  N.~Dutly, S.~Capkun, and S.~Shinde, ``{Devlore: Extending Arm CCA to
  Integrated Devices A Journey Beyond Memory to Interrupt Isolation},''
  \emph{arXiv preprint arXiv:2408.05835}, 2024.

\bibitem{barriccade}
M.~Schulze, C.~Lindenmeier, and J.~Rockl, ``{BarriCCAde: Isolating
  Closed-Source Drivers with ARM CCA},'' in \emph{2024 IEEE EuroS\&PW}, 2024.

\bibitem{aster}
M.~Kuhne, S.~Sridhara, A.~Bertschi, N.~Dutly, S.~Capkun, and S.~Shinde,
  ``{Aster: Fixing the Android TEE Ecosystem with Arm CCA},'' \emph{arXiv
  preprint arXiv:2407.16694}, 2024.

\bibitem{fortifypatch}
Z.~Ye, L.~Zhou, F.~Zhang, W.~Jin, Z.~Ning, Y.~Hu, and Z.~Qin, ``{FortifyPatch:
  Towards Tamper-Resistant Live Patching in Linux-Based Hypervisor},'' in
  \emph{{ISSTA}}, 2024.

\bibitem{cubevisor}
J.~Chen, Q.~Zhou, X.~Yan, N.~Jiang, X.~Jia, and W.~Zhang, ``{CubeVisor: A
  Multi-realm Architecture Design for Running VM with ARM CCA},'' in \emph{2024
  Annual Computer Security Applications Conference (ACSAC)}, 2024.

\bibitem{twinvisor}
D.~Li, Z.~Mi, Y.~Xia, B.~Zang, H.~Chen, and H.~Guan, ``{TwinVisor:
  Hardware-isolated Confidential Virtual Machines for ARM},'' in
  \emph{{Proceedings of the ACM SIGOPS 28th Symposium on Operating Systems
  Principles}}, 2021.

\bibitem{asguard}
M.~Moon, M.~Kim, J.~Jung, and D.~Song, ``{ASGARD}: {Protecting} {On}-{Device}
  {Deep} {Neural} {Networks} with {Virtualization}-{Based} {Trusted}
  {Execution} {Environments},'' in \emph{Proceedings 2025 {Network} and
  {Distributed} {System} {Security} {Symposium}}, 2025.

\bibitem{andrin-thesis}
A.~Bertschi,
  ``{\href{https://abertschi.ch/blog/2023-cca-trusted-peripherals/eth_mthesis_cca.pdf}{Protecting
  Accelerator Execution with Arm Confidential Computing Architecture}},''
  Master's Thesis, ETH Zurich, 2023.

\bibitem{opensgx}
P.~Jain, S.~Desai, S.~Kim, M.-W. Shih, J.~Lee, C.~Choi, Y.~Shin, T.~Kim, B.~B.
  Kang, and D.~Han, ``{OpenSGX: An Open Platform for SGX Research},'' in
  \emph{NDSS}, 2016.

\bibitem{keystone}
D.~Lee, D.~Kohlbrenner, S.~Shinde, K.~Asanovi{\'c}, and D.~Song, ``{Keystone:
  An Open Framework for Architecting Trusted Execution Environments},'' in
  \emph{EuroSys}, 2020.

\bibitem{opencca}
\BIBentryALTinterwordspacing
A.~Bertschi and S.~Shinde, ``{\href{https://opencca.github.io}{\codename: An
  Open Framework to Enable Arm CCA Research, Project Page}},'' 2025. [Online].
  Available: \url{https://opencca.github.io}
\BIBentrySTDinterwordspacing

\bibitem{collabora-rk3588}
Collabora,
  ``{\href{https://www.collabora.com/news-and-blog/news-and-events/rockchip-rk3588-upstream-support-progress-future-plans.html}{Upstream
  support for Rockchip's RK3588: Progress and future plans}},'' {Accessed:
  Feb.~1, 2025}.

\bibitem{arm-features}
ARM, ``\href{https://developer.arm.com/documentation/109697/latest}{{Feature
  names in A-profile architecture}},'' {Accessed Feb. 1, 2025}.

\bibitem{rmm-spec}
------, ``\href{https://developer.arm.com/documentation/den0137/latest/}{Realm
  Management Monitor (RMM) Specification (1.0-REL0)},'' {Accessed: Feb.~2,
  2025}.

\bibitem{uboot-binman}
U-Boot,
  ``\href{https://docs.u-boot.org/en/latest/develop/package/binman.html}{{Binman,
  Project Page}},'' {Accessed: Feb.~2, 2025}.

\bibitem{opencca-arxiv}
\BIBentryALTinterwordspacing
A.~Bertschi and S.~Shinde,
  ``{\href{https://opencca.github.io/extended-version}{\codename: An Open
  Framework to Enable Arm CCA Research, Extended Version}},'' 2025. [Online].
  Available: \url{https://opencca.github.io/extended-version}
\BIBentrySTDinterwordspacing

\bibitem{version-tfa}
Collabora,
  ``{\href{https://gitlab.collabora.com/hardware-enablement/rockchip-3588/trusted-firmware-a/-/commit/44418fce30}{\tfa:
  Upstream support for Rockchip's RK3588, Commit: 44418fce30}},'' {Accessed:
  Feb.~1, 2025}.

\bibitem{version-cca-patchset-host}
ARM,
  ``\href{https://gitlab.arm.com/linux-arm/linux-cca/-/commit/fad35572db}{CCA
  Host+Guest Patchset for Linux, Version: cca-full/v5+v7, Commit:
  fad35572db},'' {Accessed: Feb.~1, 2025}.

\bibitem{version-linux-host}
Collabora,
  ``{\href{https://gitlab.collabora.com/hardware-enablement/rockchip-3588/linux/-/commit/f7e1ed901e7}{Linux
  Kernel: Upstream support for Rockchip's RK3588, Commit: f7e1ed901e7}},''
  {Accessed: Feb.~1, 2025}.

\bibitem{version-rmm}
ARM, ``{\href{https://github.com/TF-RMM/tf-rmm/commit/1313d31ad9}{\rmm, Commit:
  1313d31ad9}},'' {Accessed: Feb. 1, 2025}.

\bibitem{version-uboot}
Collabora,
  ``{\href{https://gitlab.collabora.com/hardware-enablement/rockchip-3588/u-boot/-/commit/889c316b59e2}{\uboot:
  Upstream support for Rockchip's RK3588, Commit: 889c316b59e2}},'' {Accessed:
  Feb.~1, 2025}.

\bibitem{version-kvmtool}
ARM,
  ``{\href{https://gitlab.arm.com/linux-arm/kvmtool-cca/-/commit/54241e378}{Kvmtool
  for Arm CCA, Version: cca/v3, Commit: 54241e378}},'' {Accessed: Feb.~1,
  2025}.

\bibitem{fvp-not-cycle-accurate}
------,
  ``{\href{https://developer.arm.com/documentation/101469/2024-1/Introduction-to-Arm-Development-Studio/FVP-models?lang=en}{FVP
  Models, Cycle Accuracy}},'' {Accessed Feb. 1, 2025}.

\end{thebibliography}
